\documentclass[12pt]{iopart}

\usepackage{graphicx}

\begin{document}

\title[Polaron model revisited]{Non-linear response of molecular junctions:\\
The polaron model revisited}

\author{Michael Galperin${}^1$, Abraham Nitzan${}^2$, and Mark A. Ratner${}^3$}
\address{${}^1$ Theoretical Division and
                Center for Integrated Nanotechnologies,\\ $\ \ $
                Los Alamos National Laboratory, Los Alamos, NM 87545}
\ead{galperin@lanl.gov}
\address{${}^2$ School of Chemistry, The Sackler Faculty of Science,\\ $\ \ $
                Tel Aviv University, Tel Aviv 69978, Israel}
\ead{nitzan@post.tau.ac.il}
\address{${}^3$ Department of Chemistry and Materials Research Center,\\ $\ \ $
                Northwestern University, Evanston, IL 60208}
\ead{ratner@chem.northwestern.edu}

\begin{abstract}
A polaron model proposed as a possible mechanism for nonlinear
conductance [Galperin~M, Ratner~M~A, and Nitzan~A 2005 {\it Nano~Lett.} 
\textbf{5} 125-30] is revisited with focus on the differences
between the weak and strong molecule-lead coupling cases. Within the
one-molecular level model we present an approximate expression for
the electronic Green function corresponding to inelastic transport case, 
which in the appropriate limits reduces to expressions presented previously 
for the isolated molecule and for molecular junction coupled to 
a slow vibration (static limit). The relevance of considerations based on
the isolated molecule limit to understanding properties of molecular 
junctions is discussed. 
\end{abstract}


\submitto{\JPCM}

\section{Introduction}
Much of the interest in molecular conduction junctions stems from their 
functional properties as possible components in molecular electronic devices. 
In particular, non linear response behaviours such as bistability and 
negative differential resistance (NDR) have attracted much attention. 
Here we revisit a model for such phenomena that was previously 
advanced\cite{nanolett} and later 
criticized\cite{condmat0603467,condmat0606366,condmat0611163} 
in order to eludidate and clarify some of its mathematical characteristics.

The simplest molecular conduction junction comprises two
metallic electrodes connected by a single molecule. 
The simplest theoretical model for such a junction is a molecule 
represented by one electronic level (the molecular affinity or 
ionization level) with one vibrational mode connecting free-electron metals. 
When the molecular electronic level is outside the range between the 
lead Fermi levels and its distance from these levels 
is large compared to the strength of the molecule-lead electronic coupling, 
the transport occurs by tunneling through the molecular energy barrier. 
This is the so-called Landauer-Imry limit. When the injection gap 
(distance between the Fermi level and the affinity or ionization levels) 
becomes small, the barrier decreases, and there is an opportunity for 
stabilizing excess charge on the molecule by polarization of its
electronic and/or nuclear environment, leading to the formation 
of polaron-type trapped charge. We have previously described the 
consequences of this polarization on such phenomena as hysteresis, 
switching and negative differential resistance in molecular 
junctions.\cite{nanolett} 

When the electronic coupling between the molecule and leads vanishes, 
one deals with polaron formation on an isolated molecule, 
for which an exact solution is available. We discuss here the two limiting 
cases: polaron formation on an isolated molecule, and the transport problem 
in the limit where nuclear dynamics is slow relative to all electronic
timescales. Invoking the second case as one of the possible mechanisms of 
hysteresis, switching, and negative differential resistance in molecular 
junctions\cite{nanolett} was criticized by Alexandrov and Bratkovsky, 
in several papers.\cite{condmat0603467,condmat0606366,condmat0611163}
These authors claim that the conclusions of Ref.~\cite{nanolett}
contradict a previously published ``exact solution''\cite{AB_PRB1,AB_PRB2}
that shows no multistability is possible for molecular models comprising
nondegenerate and two-fold degenerate electronic levels.
They suggest that multistability found in Ref.~\cite{nanolett} is
``an artifact of the mean-field approximation that neglects Fermi-Dirac
statistics of electrons'' ($\hat n_0^2=\hat n_0$), 
and ``leads to a spurious self-interaction of a single polaron with 
itself and a resulting non-existent nonlinearity''. 

As was pointed out previously,\cite{reply} the weakness of this criticism
stems from using, in Ref.~\cite{AB_PRB1}, the isolated quantum dot 
limit to discuss molecular junctions. 
In contrast, we have argued\cite{reply} that the approximtion of 
Ref.~\cite{nanolett} is valid in the limit $\Gamma\gg\omega_0$,
where $\Gamma$ is the inverse lifetime of excess carrier on the bridge
and $\omega_0$ - the frequency of the relevant nuclear motion. Here we
present this argument in a rigorous mathematical form.
We describe a general
approach to this problem, which is capable reproducing the result of 
Ref.~\cite{AB_PRB1} in the isolated molecule limit
and our previous result, Ref.~\cite{nanolett}, 
in the static limit of a junction ($\omega_0/\Gamma\ll 1$),
where $\omega_0$ is the oscillator frequency and $\Gamma$,
the spectral density associated with the molecule-lead coupling,
measures the strength of this coupling. 
This validates the polaronic approach of Ref.~\cite{nanolett} in this limit.

\section{\label{neqLCE}General consideration}
One way to bridge between the limits of zero and strong molecule-lead
coupling is the nonequilibrium linked cluster expansion (NLCE)
proposed in Ref.~\cite{Kral}. 
For our purposes a first order LCE\cite{Dunn} 
(clusters of second order in electron-phonon coupling 
$M$)\footnote{We use the term ``phonon'' for any relevant molecular
or environmental vibration.} is adequate.
Indeed, this level of consideration provides exact results in both
isolated molecule and static limits, while providing an approximate expression
for the general case. The main idea of the NLCE is the same as in the usual LCE
-- one expands a Green function (GF) perturbatively in terms of the
interaction part of the Hamiltonian (in our case - the electron-phonon 
interaction) up to some finite order, 
and equates the expansion in clusters to an expression in terms of 
cumulants.\cite{Mahan}
This provides approximate resummation of the whole series.\cite{Dunn}
The NLCE considers this expansion on the Keldysh contour\cite{Kral}
\begin{equation}
 \label{NLCE}
 G(\tau,\tau') = \sum_{n=0}^{\infty}\xi^n W_n(\tau,\tau')
 = G_0(\tau,\tau')\exp\left[\sum_{n=1}^{\infty}\xi^n F_n(\tau,\tau')\right]
\end{equation}
whence, up to first order ($n=1$)
\begin{equation}
 W_0(\tau,\tau')=G_0(\tau,\tau')
 \qquad
  W_1(\tau,\tau')=G_0(\tau,\tau')F_1(\tau,\tau')
\end{equation}
Projections of (\ref{NLCE}) on the real time axis are obtained 
using Langreth rules\cite{Langreth,HaugJauho}, in particular
\begin{eqnarray}
 \label{NLCE_Gltgt}
 G^{>,<}(t,t') &=& G_0^{>,<}(t,t')
   \exp\left[\sum_{n=1}^{\infty}\xi^n F_n^{>,<}(t,t')\right]
 \\
 \label{NLCE_W0ltgt}
 W_0^{>,<}(t,t') &=& G_0^{>,<}(t,t')
 \\
 \label{NLCE_W1ltgt}
 W_1^{>,<}(t,t') &=& G_0^{>,<}(t,t')\, F_1^{>,<}(t,t')
\end{eqnarray}
In steady state (which we consider below) projections depend
on time difference $t-t'$ only.

\section{Model}
As in Ref.~\cite{nanolett} we consider a single (nondegenerate) 
electron level $\varepsilon_0$ coupled to one vibration $\omega_0$ 
and to two leads $L$ and $R$ represented by reservoirs of free electrons,
each in its own equilibrium. The vibration is represented by
a free oscillator at thermal equilibrium. 
The Hamiltonian of the system is (here and below $e=1$, $m=1$, and $\hbar=1$)
\begin{eqnarray}
 \label{H}
 \hat H &=& \varepsilon_0\hat d^\dagger\hat d 
         +  \sum_{k\in\{L,R\}}\left(\varepsilon_k\hat c_k^\dagger\hat c_k
            + V_k\hat d^\dagger\hat c_k + V_k^{*}\hat c_k^\dagger\hat d\right) 
 \nonumber \\
        &+& \omega_0\hat a^\dagger\hat a 
         +  M(\hat a+\hat a^\dagger) \hat d^\dagger \hat d
\end{eqnarray}
where $\hat d$ ($\hat d^\dagger$) and $\hat c_k$ ($\hat c_k^\dagger$)
are annihilation (creation) operators for electrons on the molecule 
and in the contacts respectively, while $\hat a$ ($\hat a^\dagger$) 
are annihilation (creation) operators of a vibrational quantum. 
The first and second terms in (\ref{H}) represent electrons on the 
bridge and in the contacts, respectively and the third and fourth terms 
describe molecule-leads coupling. The fifth term describes the free vibration,
while the last is the linear electron-phonon coupling.
For future reference we also define the operator of
molecular level population 
\begin{equation}
 \label{hatn0}
 \hat n_0=\hat d^\dagger\hat d 
\end{equation}
and its quantum and statistical average 
\begin{equation}
 \label{n0}
 n_0=<\hat n_0>=-i\int_{-\infty}^{+\infty}\frac{dE}{2\pi}G^{<}(E)
 \equiv -iG^{<}(t=0)
\end{equation}
where $G^{<}$ is the electron lesser GF.\cite{Mahan,HaugJauho}

\section{Mathematical evaluation of transport properties}
The non-equilibrium Green function technique provides a convenient framework 
for evaluating the desired transport properties. To obtain the 
steady-state current under given bias conditions
\begin{equation}
 I_K = \frac{e}{\hbar}\int_{-\infty}^{+\infty}dt\,
 \mbox{Tr}\left[\Sigma_K^{<}(-t)\,G^{>}(t) 
               -\Sigma_K^{>}(-t)\,G^{<}(t)\right]
\end{equation}
($K=L,R$) one needs to evaluate the molecular electronic Green function
in the presence of the moleule-lead and electron-phonon couplings. 
In what follows we derive this expression within the low-order 
NLCE described in Section~\ref{neqLCE}.

The free phonon GFs (retarded, advanced, lesser and greater) are
\begin{eqnarray}
 \label{D0r}
 D_0^{r}(t) &=& -i\theta(t)\left[e^{-i\omega_0 t}-e^{i\omega_0 t}\right] 
 \\
 \label{D0a}
 D_0^{a}(t) &=& i\theta(-t)\left[e^{-i\omega_0 t}-e^{i\omega_0 t}\right]
 \\
 \label{D0lt}
 D_0^{<}(t) &=& -i\left[(1+N_0)e^{i\omega_0 t}+N_0e^{-i\omega_0 t}\right]
 \\
 \label{D0gt}
 D_0^{>}(t) &=& -i\left[(1+N_0)e^{-i\omega_0 t}+N_0e^{i\omega_0 t}\right]
\end{eqnarray}
where $N_0=[e^{\omega_0/T}-1]^{-1}$ is the thermal equilibrium 
vibration population.

In the absence of electron-phonon coupling, $M=0$, electron GFs
in the wide band approximation (where the spectral densities
$\Gamma_K=2\pi\sum_{k\in K}|V_k|^2\delta(E-\varepsilon_k)$ are energy
independent) are
\begin{eqnarray}
 \label{G0r}
 G_0^{r}(t) &=& -i\theta(t)\, 
               \exp\left[-i\varepsilon_0 t-\frac{\Gamma}{2}t\right]
 \\
 \label{G0a}
 G_0^{a}(t) &=& i\theta(-t)
               \exp\left[-i\varepsilon_0 t+\frac{\Gamma}{2}t\right]
 \\
 \label{G0lt}
 G_0^{<}(t) &=& i\int_{-\infty}^{+\infty}\frac{dE}{2\pi}\, e^{-iEt}
 \frac{\Gamma_Lf_L(E)+\Gamma_Rf_R(E)}{(E-\varepsilon_0)^2+(\Gamma/2)^2}
 \nonumber \\
 &\approx & i n_0 \exp\left[-i\varepsilon_0 t-\frac{\Gamma}{2}|t|\right]
 \\
 \label{G0gt}
 G_0^{>}(t) &=& -i\int_{-\infty}^{+\infty}\frac{dE}{2\pi}\, e^{-iEt}
 \frac{\Gamma_L[1-f_L(E)]+\Gamma_R[1-f_R(E)]}{(E-\varepsilon_0)^2+(\Gamma/2)^2} 
 \nonumber \\
 &\approx & -i [1-n_0] \exp\left[-i\varepsilon_0 t-\frac{\Gamma}{2}|t|\right]
\end{eqnarray}
$\Gamma_K$ ($K=L,R$) are the electron escape rates from the molecule
due to coupling to left and right leads, $\Gamma=\Gamma_L+\Gamma_R$
and $f_K(E)=[e^{(E-\mu_k)/T}+1]^{-1}$ is the Fermi-Dirac distribution
in the contact $K$ ($\mu_K$ is chemical potential).
In approximations made in (\ref{G0lt}) and (\ref{G0gt}) we have used
$n_0\approx\left(\Gamma_Lf_L(\varepsilon_0)+\Gamma_Rf_R(\varepsilon_0)\right)
/\Gamma$.  Note that these approximations are used for
convenience only and do not influence the generality of the
considerations below. They become exact either in the case of molecule
weakly coupled to contacts or when molecular level is far 
(compared to $\Gamma$) from the contacts' chemical potentials.

\begin{figure}[htbp]
\centering\includegraphics[width=\linewidth]{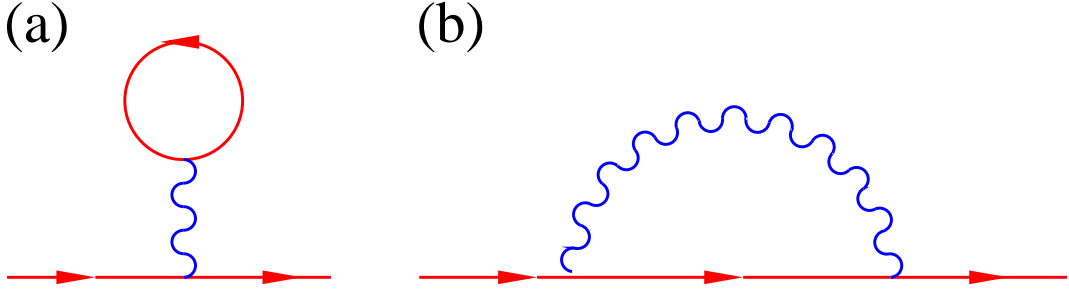}
\caption{\label{fig_diag}
Lowest ($M^2$) order contributions to electron GF ($W_1$):
(a) Hartree and (b) Fock (Born) terms. 
The wavy line represents free phonon GF. 
The straight line represents  electron GF.}
\end{figure}

The lowest order in the electron-phonon coupling ($M^2$) contribution 
to the electronic GF is given by
\begin{eqnarray}
 \label{W1_Keldysh}
 W_1(\tau,\tau') &=& \int_c d\tau_1\int_c d\tau_2\,
 G_0(\tau,\tau_1)\,\Sigma_{ph}(\tau_1,\tau_2)\, G_0(\tau_2,\tau')
 \\
 \label{Sigma}
 \Sigma_{ph}(\tau_1,\tau_2) &=& 
    \delta(\tau_1,\tau_2)M^2 n_0\int_c d\tau_3 D_0(\tau_1,\tau_3)
    + i M^2 D_0(\tau_1,\tau_2)\, G_0(\tau_1,\tau_2)
\end{eqnarray}
where self-energy (SE) $\Sigma_{ph}$ is a sum of two contributions:
the first and second terms in Eq.(\ref{Sigma}) are respectively the
Hartree and Born terms shown in Fig.~\ref{fig_diag}. 
The importance of including the Hartree term 
when considering systems without translational periodicity 
(e.g. molecular junctions) was emphasized in a number of 
papers.\cite{Wilkins,Hewson,Kral} 

The lesser and greater projections of (\ref{W1_Keldysh}) onto the real 
time axis (here and below we assume steady-state situation) are 
obtained from the Langreth rules\cite{Langreth,HaugJauho} as
\begin{eqnarray}
 \label{W1ltgt}
 &&W_1^{>,<}(t) = \int_{-\infty}^{+\infty} dt_1\int_{-\infty}^{+\infty}dt_2\,
 \left[ G_0^{r}(t-t_1)\,\Sigma_{ph}^{>,<}(t_1-t_2)\, G_0^{a}(t_2)
 \right.\\ &&+ \left.
   G_0^{>,<}(t-t_1)\,\Sigma_{ph}^{a}(t_1-t_2)\, G_0^{a}(t_2)
 + G_0^{r}(t-t_1)\,\Sigma_{ph}^{r}(t_1-t_2)\, G_0^{>,<}(t_2)
 \right]
 \nonumber
\end{eqnarray}
Projecting (\ref{Sigma}) and using Eqs.~(\ref{D0r})-(\ref{G0gt})
one gets
\begin{eqnarray}
 \label{SEr}
 \Sigma_{ph}^{r}(t) &=& -iM^2\theta(t)\left[
  (N_0+1-n_0)e^{-i\omega_0t} +(N_0+n_0)e^{i\omega_0 t}
 \right]e^{-i\varepsilon_0 t-\Gamma t/2}
 \nonumber \\
 &-& \delta(t)\frac{2M^2n_0}{\omega_0}
 \\
 \label{SEa}
 \Sigma_{ph}^{a}(t) &=& iM^2\theta(-t)\left[
  (N_0+1-n_0)e^{-i\omega_0 t}  
 +(N_0+n_0)e^{i\omega_0 t}
 \right]e^{-i\varepsilon_0 t+\Gamma t/2}
 \nonumber \\
 &-& \delta(t)\frac{2M^2n_0}{\omega_0}
 \\
 \label{SElt}
 \Sigma_{ph}^{<}(t) &=& iM^2 n_0 e^{-i\varepsilon_0 t - \Gamma |t|/2}
 \left[(1+N_0)e^{i\omega_0 t} + N_0 e^{-i\omega_0 t}\right]
 \\
 \label{SEgt}
 \Sigma_{ph}^{>}(t) &=& -iM^2 [1-n_0] e^{-i\varepsilon_0 t - \Gamma |t|/2}
 \left[(1+N_0)e^{-i\omega_0 t} + N_0 e^{i\omega_0 t}\right]
\end{eqnarray}
It should be emphasized that the $n_0$ term that enters the Hartree 
contribution in Eqs.~(\ref{Sigma}), (\ref{SEr}) and (\ref{SEa}) 
is an exact result; unrelated to the convenient approximation 
made in Eqs.~(\ref{G0lt}) and (\ref{G0gt}) above (that leads to the
explicit appearance of the $n_0$ terms in Eqs.~(\ref{SElt}) and (\ref{SEgt})).
It is this term which will provide the population dependent shift 
of the electronic level in the static limit (see below).

Our aim is to get an expression for the retarded electron GF
\begin{equation}
 \label{Gr}
 G^r(t) = \theta(t) \left[G^{>}(t)-G^{<}(t)\right]
\end{equation}
using (\ref{NLCE_Gltgt})-(\ref{NLCE_W1ltgt}).
In order to do so we have to calculate $W_1^{>,<}$ which is given by
Eq.(\ref{W1ltgt}). It is convenient to consider separately the
first term and the sum of the second and third terms on the
right-hand-side in (\ref{W1ltgt})
\begin{eqnarray}
 \label{W1ltgt_I12}
 W_1^{>,<}(t) &=& {\cal W}_1^{>,<}(t) + {\cal W}_2^{>,<}(t)
 \\
 {\cal W}_1^{>,<}(t) &=& 
 \int_{-\infty}^{+\infty} dt_1 \int_{-\infty}^{+\infty}dt_2\,
 G_0^{r}(t-t_1)\,\Sigma_{ph}^{>,<}(t_1-t_2)\, G_0^{a}(t_2)
 \\
 {\cal W}_2^{>,<}(t) &=& 
 \int_{-\infty}^{+\infty} dt_1\int_{-\infty}^{+\infty}dt_2\,
 \left[
   G_0^{>,<}(t-t_1)\,\Sigma_{ph}^{a}(t_1-t_2)\, G_0^{a}(t_2)
 \right.\nonumber \\ && \left. \qquad\qquad\qquad\quad
 + G_0^{r}(t-t_1)\,\Sigma_{ph}^{r}(t_1-t_2)\, G_0^{>,<}(t_2)
 \right]
\end{eqnarray}
utilizing (\ref{G0r})-(\ref{G0gt}) and (\ref{SEr})-(\ref{SEgt})
then leads to
\begin{eqnarray}
 \label{I1lt}
 &&{\cal W}_1^{<}(t) = i n_0 e^{-i\varepsilon_0 t-\Gamma |t|/2}
 \frac{M^2}{\omega_0}
 \\ &&\times 
 \left\{ \theta(t)
 \left[(1+N_0)\left(\frac{e^{i\omega_0 t}}{\omega_0+i\Gamma}
                   +\frac{1}{\omega_0-i\Gamma}\right)
      + N_0\left(\frac{e^{-i\omega_0 t}}{\omega_0-i\Gamma}
                   +\frac{1}{\omega_0+i\Gamma}\right)\right]
 \right\}
 \nonumber \\ && + \left. \theta(-t)
 \left[(1+N_0)\left(\frac{e^{i\omega_0 t}}{\omega_0-i\Gamma}
                   +\frac{1}{\omega_0+i\Gamma}\right)
      + N_0\left(\frac{e^{-i\omega_0 t}}{\omega_0+i\Gamma}
                   +\frac{1}{\omega_0-i\Gamma}\right)\right]
 \right\}
 \nonumber 
 \\
 \label{I2lt}
 &&{\cal W}_2^{<}(t) = - i n_0 e^{-i\varepsilon_0 t-\Gamma |t|/2}
 \frac{M^2}{\omega_0} 
 \nonumber \\ &&\times 
 \left\{ \theta(t)
 \left[(N_0+1-n_0)\left(\frac{1}{\omega_0+i\Gamma}+\frac{1}{\omega_0}-it 
       +e^{-i\omega_0 t}\left[\frac{1}{\omega_0-i\Gamma}-\frac{1}{\omega_0}
       \right]\right)
 \nonumber \right.\right. \\ &&\qquad\qquad \left.
      +(N_0+n_0)\left(\frac{1}{\omega_0-i\Gamma}+\frac{1}{\omega_0}+it
       +e^{i\omega_0 t}\left[\frac{1}{\omega_0+i\Gamma}-\frac{1}{\omega_0}
       \right]\right)\right]
 \\ && + \theta(-t)
 \left[(N_0+1-n_0)\left(\frac{1}{\omega_0-i\Gamma}+\frac{1}{\omega_0}-it
       +e^{-i\omega_0 t}\left[\frac{1}{\omega_0+i\Gamma}-\frac{1}{\omega_0} 
       \right]\right)
 \right. \nonumber \\ &&\qquad\qquad \left.
      +(N_0+n_0)\left(\frac{1}{\omega_0+i\Gamma}+\frac{1}{\omega_0}+it
       +e^{i\omega_0 t}\left[\frac{1}{\omega_0-i\Gamma}-\frac{1}{\omega_0}      
  \right]\right)\right]
 \nonumber \\ &&\quad - \left.
  2 i n_0 t \right\}
 \nonumber
\end{eqnarray}
The last term in curly brackets on the right-hand-side in (\ref{I2lt}) 
comes from the Hartree term.
The expression for ${\cal W}_1^{>}(t)$ is obtained from (\ref{I1lt})
by interchanging $N_0$ and $N_0+1$ and replacing $n_0$ by $n_0-1$.
${\cal W}_2^{>}(t)$ is obtained from (\ref{I2lt})
by replacing $n_0$ by $n_0-1$ only in the prefactor that multiplies
the curly brackets on the right-hand-side.
These general approximate (first order LCE) expressions 
for ${\cal W}^{>,<}_{1,2}$ are the central result of this consideration.

\section{Two physical limits}
In \cite{nanolett} we have discussed a mean field approach to 
describe non-linear response of molecular junctions characterised by 
strong molecule-lead coupling as well as slow vibrations strongly coupled 
to the electronic subsystem. As noted in the introduction this approach 
was criticised in 
Refs.~\cite{condmat0603467,condmat0606366,condmat0611163}
as incompatible with observations made in the isolated molecule. 
To elucidate the issue we consider next these two specific limits: 
the isolated molecule ($\Gamma\to 0$) and static limit 
($\omega_0/\Gamma\to 0$).

\subsection*{The isolated molecule}
In the limit $\Gamma\to 0$ Eqs.~(\ref{I1lt}) and (\ref{I2lt}) yield
\begin{eqnarray}
 \label{I1lt_Gamma_0}
 {\cal W}_1^{<}(t) &=& i n_0 e^{-i\varepsilon_0 t} \frac{M^2}{\omega_0^2} 
 \nonumber \\ &\times &
 \left[(2N_0+1)+(1+N_0)e^{i\omega_0 t}+N_0e^{-i\omega_0 t}\right]
 \\
 \label{I2lt_Gamma_0}
 {\cal W}_2^{<}(t) &=&  i n_0 e^{-i\varepsilon_0 t} 
 \left[-2\frac{M^2}{\omega_0^2}(2N_0+1)+i\frac{M^2}{\omega_0}t\right]
\end{eqnarray}
and the corresponding expressions for ${\cal W}_{1,2}^{>}(t)$
\begin{eqnarray}
 \label{I1gt_Gamma_0}
 {\cal W}_1^{>}(t) &=& -i[1-n_0] e^{-i\varepsilon_0 t} \frac{M^2}{\omega_0^2} 
 \nonumber \\ &\times &
 \left[(2N_0+1)+(1+N_0)e^{-i\omega_0 t}+N_0e^{i\omega_0 t}\right]
 \\
 \label{I2gt_Gamma_0}
 {\cal W}_2^{>}(t) &=&  -i[1-n_0] e^{-i\varepsilon_0 t} 
 \left[-2\frac{M^2}{\omega_0^2}(2N_0+1)+i\frac{M^2}{\omega_0}t\right]
\end{eqnarray}
Substituting (\ref{I1lt_Gamma_0})-(\ref{I2gt_Gamma_0}) into (\ref{W1ltgt_I12}) 
and using Eqs.~(\ref{NLCE_Gltgt}) and (\ref{NLCE_W1ltgt}), 
one gets from (\ref{Gr})
\begin{eqnarray}
 \label{Gr_Gamma_0}
 G^{r}(t) &=& -i\theta(t) e^{-i(\varepsilon_0-\Delta)t}e^{-\lambda^2(2N_0+1)}
 \nonumber \\
 &\times &\left\{
  (1-n_0)\exp\left(\lambda^2
         \left[N_0e^{i\omega_0 t}+(1+N_0)e^{-i\omega_0 t}\right]\right)
 \right. \\ &+& \left.
 \qquad n_0\quad\,\,   \exp\left(\lambda^2
         \left[N_0e^{-i\omega_0 t}+(1+N_0)e^{i\omega_0 t}\right]\right)
 \right\}
 \nonumber
\end{eqnarray}
where
\begin{equation}
 \Delta\equiv\frac{M^2}{\omega_0} \qquad \lambda\equiv\frac{M}{\omega_0}
\end{equation}
Eq.(\ref{Gr_Gamma_0}) is the standard expression for the retarded Green function
in the isolated molecule case, obtained following a small polaron 
(Lang-Firsov or canonical) transformation.\cite{Mahan}
In particular, it is identical to Eq.(30) of Ref.~\cite{AB_PRB1} for the
case of a nondegenerate level (i.e. $d=1$ there). 
Note that approximations (\ref{G0lt}) and (\ref{G0gt})
become exact in this limit and, furthermore,
the first order LCE provides the exact result in this limit.
As was pointed out by Alexandrov and 
Bratkovsky\cite{condmat0603467,condmat0606366,condmat0611163}
the electronic level shift, $\Delta$,
is independent of level population for the isolated molecule,
and no multistability is possible in this case.

\subsection*{The static limit}
The $\omega_0/\Gamma\to 0$ limit reflects either a slow vibration or a 
strong molecule-lead coupling. For molecules chemisorbed on metal and 
semiconductor surfaces $\Gamma$ is often of order $0.1-1$~eV, 
so this limit is expected to be relevant for the relatively slow molecular 
motions associated with molecular configuration changes. 
To describe the behaviour of our model system in this case we expand 
the exponentials and the fractions in Eqs.~(\ref{I1lt}) and (\ref{I2lt}) 
in powers of $\omega_0/\Gamma$, disregarding terms of order higher than 
$1$ and keeping in mind that due to the $e^{-\Gamma|t|/2}$ prefactor 
$\omega_0/\Gamma\sim\omega_0 t$ holds. This implies
\begin{equation}
 e^{\pm i\omega_0 t} \approx 1 \pm i\omega_0 t
 \qquad
 \frac{1}{\omega_0\pm i\Gamma} \approx 
 \frac{1}{\pm i\Gamma}\left(1\mp \frac{\omega_0}{i\Gamma}\right)
\end{equation}
which leads to
\begin{eqnarray}
 \label{I1lt_w0_0}
 {\cal W}_1^{<}(t) &=& 0
 \\
 \label{I2lt_w0_0}
 {\cal W}_2^{<}(t) &=& i n_0 e^{-i\varepsilon_0 t-\Gamma |t|/2} 
 \, i\frac{2M^2n_0}{\omega_0}t
\end{eqnarray}
and corresponding expressions for ${\cal W}_{1,2}^{>}(t)$
\begin{eqnarray}
 \label{I1gt_w0_0}
 {\cal W}_1^{>}(t) &=& 0
 \\
 \label{I2gt_w0_0}
 {\cal W}_2^{>}(t) &=& -i [1-n_0] e^{-i\varepsilon_0 t-\Gamma |t|/2} 
 \, i\frac{2M^2n_0}{\omega_0}t
\end{eqnarray}
Substituting (\ref{I1lt_w0_0})-(\ref{I2gt_w0_0}) into (\ref{W1ltgt_I12}) 
and using the result in (\ref{NLCE_Gltgt}) and (\ref{NLCE_W1ltgt}), 
one gets from (\ref{Gr})
\begin{equation}
 \label{Gr_w0_0}
 G^r(t) = -i\theta(t) e^{-i(\varepsilon_0-2n_0\Delta)t-\Gamma t/2}
\end{equation}
Again we note that the factor $n_0$ that enters this expression 
does not result from approximations (\ref{G0lt}) and (\ref{G0gt}). 
Rather, it arises from the exact expression for the Hartree term,
the first term on the right-hand-side in Eq.(\ref{Sigma}).
Note also, that the approximation used in Eqs.~(\ref{G0lt}) and (\ref{G0gt})
could in principle be relaxed. This would make the mathematical
evaluation more difficult (unless the molecular level is far,
compared to $\Gamma$, from the leads' chemical potentials, when this
approximation becomes exact) but would not influence the estimates 
of ${\cal W}_{1,2}^{>,<}$ in terms of $\omega_0/\Gamma$.

Note that technically the static limit corresponds to disregarding all 
diagrams (in all orders of electron-phonon interaction) except the 
Hartree term (see Fig.~\ref{fig_diag}a) and terms of similar character
(only diagrams with boson lines terminated in a closed loop), 
since these are the only diagrams transmitting zero frequency. 
In the static limit this is not a mean-field approximation but an exact 
result. Detailed discussion on the issue can be found in 
Ref.~\cite{Hewson}.

To conclude, in the static limit (which is the limit considered in 
Ref.~\cite{nanolett}) the electronic level shift, $2n_0\Delta$, 
does depend on level population
in the way presented in our polaron model.\cite{nanolett}
In what follows we briefly reiterate the implications of this
observation on the conduction properties of molecular junctions
with strong coupling between the electronic and nuclear 
subsystem.\cite{nanolett}

\section{Non-linear conduction in static limit}
Here we discuss briefly the consequences of the reorganization energy
dependence on the average electronic population in the molecule, 
as presented in Eq.(\ref{Gr_w0_0}), on the junction transport
properties.
Since we consider steady-state transport, i.e. all
GFs and SEs depend on time difference only, we can go to the energy domain.
The Fourier transform of Eq.(\ref{Gr_w0_0}) is
\begin{equation}
 G^r(E) = \left[E-\tilde\varepsilon_0(n_0)+i\Gamma/2\right]
\end{equation}
where $\tilde\varepsilon_0(n_0)\equiv\varepsilon_0-2n_0\Delta$ 
is the population dependent energy of the molecular level. 
Using the Keldysh equation
\begin{equation}
 G^{>,<}(E) = G^r(E)\,\Sigma^{>,<}(E)\, G^a(E)
\end{equation}
in expression (\ref{n0}) for the level population leads to
\begin{equation}
 \label{n0_sc}
 n_0 = \int_{-\infty}^{+\infty}\frac{dE}{2\pi}\,
 \frac{f_L(E)\Gamma_L+f_R(E)\Gamma_R}
      {[E-\tilde\varepsilon_0(n_0)]^2+[\Gamma/2]^2}
\end{equation}
This is the central result of Ref.~\cite{nanolett} (see Eq.(13) there).
The non-linear character of Eq.(\ref{n0_sc}) with respect to $n_0$ 
leads to possibility of multistability, 
and results in non-linear character of the junction transport.
In particular, the zero-temperature case allows analytical evaluation of
the integral. We find that Eq.(\ref{n0_sc}) is equivalent to the following 
pair of equations (see Eq.(20) of Ref.~\cite{nanolett})
\begin{equation}
 \label{n0_sc_0K}
 \left\{
 \begin{array}{rcl}
  n_0 & = & 
  \frac{\Gamma_L}{\pi\Gamma}\arctan\left(x+\frac{2\Gamma_R V}{\Gamma^2}\right)
 +\frac{\Gamma_R}{\pi\Gamma}\arctan\left(x-\frac{2\Gamma_L V}{\Gamma^2}\right)
 +\frac{1}{2}
 \\
 n_0 & = & \frac{\Gamma}{4\Delta}\, x +\frac{\varepsilon_0-E_F}{2\Delta} 
 \end{array}
 \right.
\end{equation}
where  $V$ is source-drain voltage. System of equations (\ref{n0_sc_0K}) 
defines points of intersection of an $\arctan$ function with a straight line,
which for some set of parameters may have multiple solutions.
Detailed discussion of the consequences of this multistability 
for transport can be found in Ref.~\cite{nanolett}
   
\section{Conclusion}
In this paper we have presented solid theoretical foundations for 
the polaron model of non-linear response of molecular junctions, 
which was proviously introduced using mean field arguments.
We have used the non-equilibrium linked cluster expansion to second order, 
and focused on the limit of the isolated molecular polaron on one hand, 
and the polaron formation in a functioning molecular transport junction 
(that is, with finite coupling to the electronic states in the leads)
on the other. 
Proper examination shows that the former case indeed requires integral 
charge on the molecule (this is self evident, since there is no source or 
drain for the electrons). The functioning junction can have a non-integer
average population of electrons on the molecule, and is 
maintained at steady state by the actual current flow through the molecule. 

This formal analysis demonstrates the validity of the polaron model as 
originally suggested, and shows clearly an example of a new
molecular regime of functioning transport junctions, characterized by 
strong molecule-lead coupling and slow molecular vibrations strongly 
coupled to the electronic population on the molecule, where the junction 
effect on its environment can be described by its {\em non-integral}\/ 
electronic population. 
Furthermore it shows that in this case, due to the phonon 
polarization, the electronic level energy becomes dependent on this 
population. This is not a ``spurious self-interaction'' (as suggested in 
Refs.~\cite{condmat0603467,condmat0606366,condmat0611163}), 
but rather describes the interaction of a tunneling electron with its 
predecessor(s) via the phonon polarization cloud created by the 
electronic transient density of the molecule.

Finally, while we believe the mathematical issues concerning the model 
advanced in Ref.~\cite{nanolett} has now been clarified, 
it should be pointed out that actual observations of multistability and 
NDR in molecular junctions can arise from other mechanisms. 
In particular, to account for such observations in the Coulomb blockade 
regime we would probably need to go beyond the simple model considered here.

\ack
M.G. thanks Yuri M. Galperin, Ivar Martin, and Andrei Komnik for stimulating
discussions.  M.G. gratefully acknowledges the support of a LANL Director's 
Postdoctoral Fellowship. 
A.N. thanks the Israel Science Foundation, the US-Israel Binational 
Science Foundation and the German-Israel Foundation for financial support.
M.R. thanks the NSF/MRSEC for support, through the NU-MRSEC.

\end{document}